\documentclass[10pt,conference]{IEEEtran}
\IEEEoverridecommandlockouts

\usepackage{amsmath,amsfonts}
\usepackage{graphicx}
\usepackage{textcomp}
\usepackage{enumitem}
\usepackage{multirow}
\usepackage{colortbl}
\usepackage{array}
\usepackage{tcolorbox}
\usepackage{colortbl}
\usepackage{nicematrix}
\usepackage{tabularx}
\usepackage{balance}
\usepackage{colortbl}
\usepackage{booktabs}

\usepackage{siunitx}
\usepackage{makecell}

\newcommand{\pval}[2]{%
  \ifnum#1<50 \textbf{\small (#2)} \else \small (#2) \fi
}

\definecolor{lightgray}{gray}{0.9}

\PassOptionsToPackage{normalem}{ulem}
\usepackage{ulem}

\providecolor{added}{rgb}{0,0,1}
\providecolor{deleted}{rgb}{1,0,0}

\usepackage{subcaption}
\usepackage{mwe}

\definecolor{specialblue}{RGB}{0,104,149}

\definecolor{specialgray}{RGB}{242,242,242}

\usepackage{gb4e}
\noautomath

\usepackage{tikz}
\usepackage[framemethod=tikz]{mdframed}
\usetikzlibrary{calc}

\newtcolorbox{mybox}{
  sharp corners,
  colback=specialgray,
  colframe=specialblue,
  boxrule=0pt,
    toprule=0pt,
  bottomrule=0pt,
  leftrule=3pt, 
  rightrule=3pt 
}

\usepackage{cite}
\usepackage{amsmath,amssymb,amsfonts}
\usepackage{algorithmic}
\usepackage{graphicx}
\usepackage{textcomp}
\def\BibTeX{{\rm B\kern-.05em{\sc i\kern-.025em b}\kern-.08em
    T\kern-.1667em\lower.7ex\hbox{E}\kern-.125emX}}
\begin{document}

\title{Zero-Shot Prompting Approaches for LLM-based Graphical User Interface Generation}

\author{
\IEEEauthorblockN{
Kristian Kolthoff\IEEEauthorrefmark{2}\IEEEauthorrefmark{1}, 
Felix Kretzer\IEEEauthorrefmark{3}\IEEEauthorrefmark{1}, 
Lennart Fiebig\IEEEauthorrefmark{2}, 
Christian Bartelt\IEEEauthorrefmark{2}, \\
Alexander Maedche\IEEEauthorrefmark{3},
and Simone Paolo Ponzetto\IEEEauthorrefmark{4}}

\IEEEauthorblockA{\IEEEauthorrefmark{1}
Authors contributed equally to the paper.}

\IEEEauthorblockA{
\IEEEauthorrefmark{2}Institute for Enterprise Systems,
University of Mannheim, Mannheim, Germany\\
Email: \{kolthoff, bartelt\} @es.uni-mannheim.de, fiebiglennart@gmail.com}

\IEEEauthorblockA{\IEEEauthorrefmark{3}
Human-Centered Systems Lab,
Karlsruhe Institute of Technology, Karlsruhe, Germany\\
Email: \{felix.kretzer, alexander.maedche\} @kit.edu}

\IEEEauthorblockA{\IEEEauthorrefmark{4}
Data and Web Science Group, 
University of Mannheim, Mannheim, Germany\\
Email: simone@informatik.uni-mannheim.de}}

\maketitle

\begin{abstract}
Graphical user interface (GUI) prototyping represents an essential activity in the development of interactive systems, which are omnipresent today. GUI prototypes facilitate elicitation of requirements and help to test, evaluate, and validate ideas with users and the development team. However, creating GUI prototypes is a time-consuming process and often requires extensive resources. While existing research for automatic GUI generation focused largely on resource-intensive training and fine-tuning of LLMs, mainly for low-fidelity GUIs, we investigate the potential and effectiveness of Zero-Shot (ZS) prompting for high-fidelity GUI generation. We propose a Retrieval-Augmented GUI Generation (RAGG) approach, integrated with an LLM-based GUI retrieval re-ranking and filtering mechanism based on a large-scale GUI repository. In addition, we adapt Prompt Decomposition (PDGG) and Self-Critique (SCGG) for GUI generation. To evaluate the effectiveness of the proposed ZS prompting approaches for GUI generation, we extensively evaluated the accuracy and subjective satisfaction of the generated GUI prototypes. Our evaluation, which encompasses over 3,000 GUI annotations from over 100 crowd-workers with UI/UX experience, shows that SCGG, in contrast to PDGG and RAGG, can lead to more effective GUI generation, and provides valuable insights into the defects that are produced by the LLMs in the generated GUI prototypes.
\end{abstract}


\IEEEpeerreviewmaketitle

\section{Introduction}

Graphical user interfaces (GUIs) enable users to easily interact with software systems, therefore representing an important artifact during the development of interactive systems. Moreover, GUI prototypes are versatile in their application scenarios and can be constructed for various purposes with different requirements. For example, GUI prototyping is a crucial activity during the early stages of software development in the requirements elicitation phase facilitating the clarification and continuous communication of requirements with stakeholders via tangible artifacts \cite{rudd1996low, baumer1996user, BRHEL2015163}. Providing GUI prototypes to customers integrates them early into the development activities and can spark meaningful discussions with requirements analysts to avoid misunderstandings. In particular, high-fidelity GUI prototypes enable the gathering of more specific feedback \cite{rudd1996low, coyette2007multi}. In addition, GUI prototypes can serve GUI designers to rapidly generate various design ideas following a parallel prototyping approach enabling the assessment of the potential of alternative GUI designs at an early stage \cite{baumer1996user}. Consequently, the usage of GUI prototypes can therefore contribute to a final software product of higher quality. However, the prototyping of high-fidelity GUIs simultaneously carries the shortcoming that the creation of these GUI prototypes is expensive and time-consuming \cite{rudd1996low}. In addition, while GUI prototypes being employed for requirements elicitation, constant change and refining of requirements is typical \cite{debnath2021ideas}, necessitating frequent GUI prototype redesign that requires further resources.

To mitigate these shortcomings, recent research focused on automatically generating GUI prototypes. For example, \textit{Instigator} \cite{brie2023evaluating} is based on \textit{minGPT} \cite{minGPT} trained on vast amounts of crawled web pages to produce low-fidelity GUI layouts based on text descriptions and user-selected GUI component types. An alternative to training a task-specific GPT model from scratch involves fine-tuning a pretrained LLM, leveraging the \textit{Rico} \cite{deka2017rico} GUI repository \cite{feng2023designing}. However, both approaches not only require resource-intensive training, but also suffer from being able to merely create low-fidelity layouts in a domain-specific language (DSL), which is hard to integrate in practical GUI prototyping workflows. Furthermore, \textit{MAxPrototyper} \cite{yuan2024maxprototyper} enables the creation of high-fidelity GUI prototypes by prompting LLMs. However, their approach requires not only text descriptions as input, but also a fully developed GUI layout and generates a proprietary DSL. In addition, their approach mainly creates relevant content (e.g., text and images) but neglects creating actual GUI prototype functionality. Recent work on employing zero-shot prompting for GUI prototyping focused solely on the verification of requirements implementation in GUI prototypes \cite{kolthoff2024interlinking}, however, misses to investigate zero-shot prompting for generating entire GUI prototypes from descriptions.

While existing research has focused solely on resource-intensive training and fine-tuning to generate low-fidelity GUI prototypes, a comprehensive investigation into less resource-intensive approaches building on zero-shot prompting techniques for generating high-fidelity GUI prototypes is currently lacking. In order to address this research gap, in this work, we explore the potential and effectiveness of different ZS prompting approaches for generating high-fidelity GUI prototypes from short high-level text descriptions (NLR) in HTML/CSS. Our focus lies particularly in ZS prompting due to \textit{(i)} training or fine-tuning of LLMs requires resource-intensive training, and \textit{(ii)} few-shot prompting struggles with very large context windows for In-Context Learning (ICL) \cite{dong2022survey, li2024long}, which would be necessary for GUI generation, since high-quality GUIs typically consist of thousands of tokens. In particular, we propose Retrieval-Augmented GUI Generation (RAGG), which combines the advantages of GUI retrieval approaches in terms of the rapid access to the vast prototyping knowledge embodied in large GUI repositories with the reasoning and adaption capabilities of LLMs. To improve the relevance of the retrieved GUI prototypes, we propose an LLM-based re-ranking approach that significantly outperforms state-of-the-art (SOTA) approaches \cite{kolthoff2023data}. Moreover, we investigate ZS Prompt Decomposition \cite{khot2022decomposed} for GUI generation (PDGG), enabling the LLM to generate meaningful intermediate reasoning outputs, instead of directly generating low-level HTML/CSS from high-level NLR. This approach more closely follows a human expert process and ensures that computation capabilities of the LLM are more sufficiently used. Finally, we investigate Self-Critique \cite{saunders2022self} for GUI generation (SCGG), employing the LLM itself in a GUI prototyping and feedback loop. Our source code, datasets and additional material is publicly available in our repository \cite{zs-prompting-github}.

To assess our approach, we conducted an extensive evaluation consisting of over 3,000 GUI annotations from over 100 crowd-workers with UI/UX experienc regarding the accuracy and subjective satisfaction of generated GUIs. The results indicate that the proposed ZS prompting approaches can significantly enhance GUI prototypes in many aspects compared to ZS baselines.

\noindent To summarize, we make the following main contributions:

\begin{itemize}[leftmargin=0.15in]
    \item We propose an LLM-based GUI re-ranking mechanism that significantly outperforms SOTA approaches
    \item We propose a Retrieval-Augmented GUI Generation (RAGG) approach leveraging a large-scale GUI repository and adapt Prompt Decomposition (PDGG) and Self-Critique (SCGG) ZS prompting for GUI generation
    \item We conduct an extensive evaluation by obtaining more than 3,000 GUI annotations from over 100 skilled crowd-workers and provide insights into defects of LLM-generated GUIs
\end{itemize}
\section{Background}

In this section, we briefly summarize important research that we build upon, including NL-based GUI retrieval and LLMs.

\subsection{NL-based GUI Retrieval}

To achieve NL-based GUI retrieval for our proposed RAGG method, we closely follow the RaWi \cite{kolthoff2023data} approach. This GUI retrieval approach initially employs a simpler model that can be computed rapidly over the \textit{Rico} GUI repository \cite{deka2017rico}, a large-scale publicly available dataset including GUI screenshots and hierarchy data of mobile apps. For example, \textit{TF-IDF} based \textit{BM25} \cite{robertson1995okapi} or a neural embedding based \textit{SentenceBERT} model using cosine similarity are employed. We combine their method with the idea of using \textit{Screen2Words (S2W)} \cite{wang2021screen2words} proposed in \textit{MAxPrototyper} \cite{yuan2024maxprototyper}, to match NLR with high-level GUI screenshots descriptions using an embedding model.

\subsection{Large Language Models (LLMs)}

LLMs are large-scale generative models with billions of parameters pretrained on vast amounts of textual data based on the transformer architecture \cite{vaswani2017attention}. LLMs especially gained popularity with the release of GPT-3 \cite{brown2020language}, showing that LLMs posses not only few-shot learning capabilities, but also are zero-shot learners \cite{radford2019language} which allows for rapid adaption to novel tasks where the models have not specifically been trained on. A plethora of effective prompting approaches has been proposed in research before \cite{liu2023pre}, ranging from simple instruction-based ZS prompting \cite{radford2019language}, Chain-of-Thought (CoT) prompting \cite{kojima2022large, wei2022chain}, Prompt Decomposition \cite{khot2022decomposed}, Self-Consistency in CoT \cite{wang2022self}, Self-Critique \cite{saunders2022self} to more sophisticated approaches such as Tree-of-Thoughts (ToT) \cite{yao2024tree, long2023large}, Graph-of-Thoughts (GoT) \cite{besta2024graph, yao2023beyond}, Graph-Chain-of-Thought (GCoT) \cite{jin2024graph}, Retrieval-Augmented Generation (RAG) \cite{li2022survey, gao2023retrieval, wu2024retrieval}. While these prompting approaches show promising potential, they have not been investigated for GUI generation.

\section{Methodology}

\begin{figure*}
  \centering
 \includegraphics[width=0.95\textwidth]{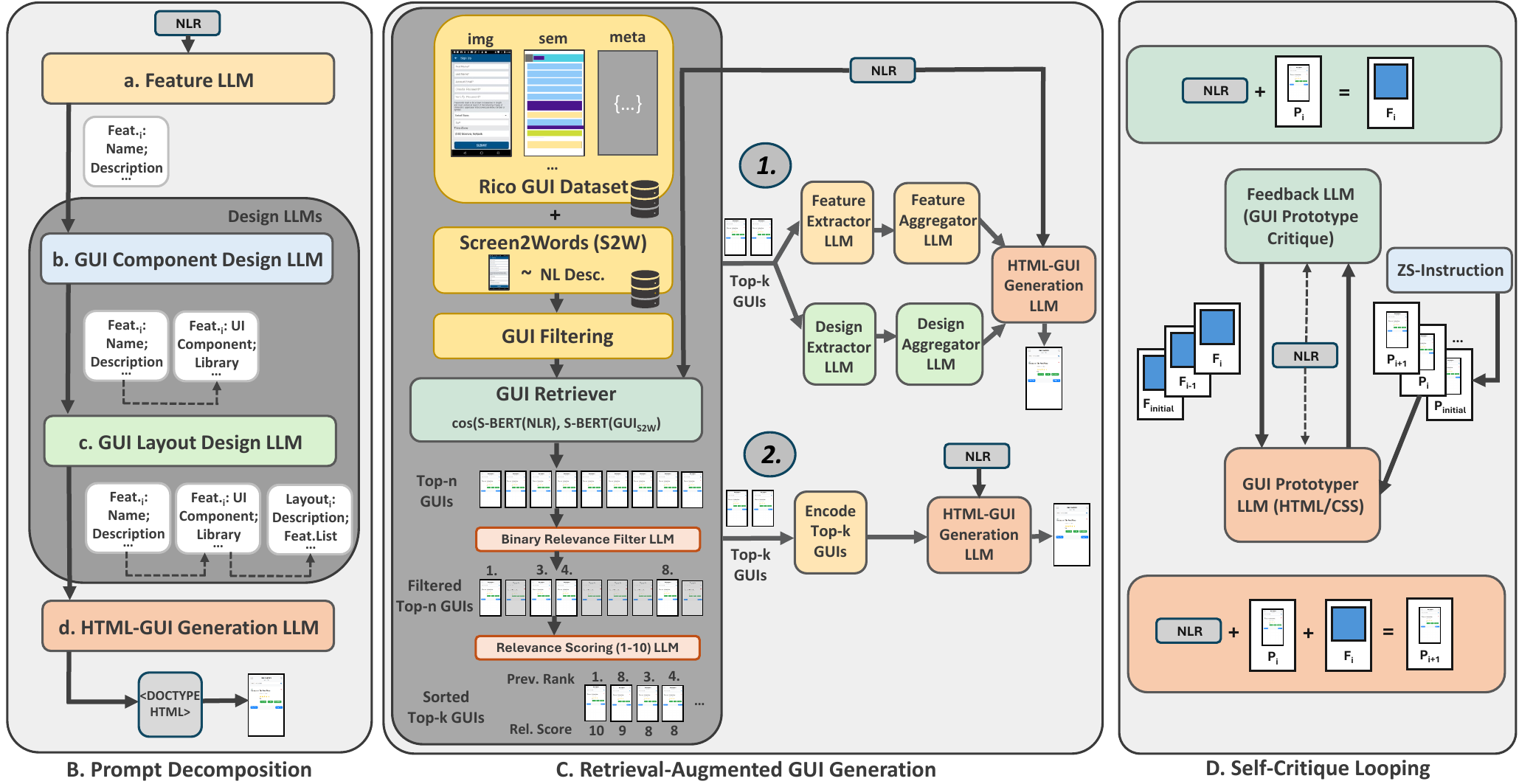}
  \caption{Overview of our ZS prompting approaches for generating GUI prototypes from natural language requirements (NLR)}
	\label{fig:overview}
\end{figure*}

In this section, we present the different ZS prompting methods for generating GUI prototypes. First, we introduce the ZS baselines employed to benchmark the more comprehensive approaches. Subsequently, we present our combined Prompt Decomposition (PDGG), RAGG and Self-Critique (SCGG) prompting approaches for GUI generation (overview Fig. \ref{fig:overview}).

\subsection{Baselines: ZS-Instruction and ZS-CoT}

As our ZS prompting baselines, we first employ a ZS instruction prompt, including a clear description of the base task to create a mobile page in HTML/CSS according to the provided brief textual description. As the target language we decided to employ HTML/CSS, since it is a widely used description language for GUIs, enables interaction with the GUI and the LLMs are typically pretrained on this format, leveraging the LLMs capabilities. To improve the alignment of LLM responses for user interaction, LLMs are often optimized with reinforcement learning with human feedback \cite{ouyang2022training}, leading to providing explanations and a structured representation of the response. In particular, the LLMs for GUI generation have a tendency to provide explanations and separate the HTML/CSS in markdown blocks. To avoid this, we additionally instruct the LLM to directly output the non-separated code without explanation. In addition, we employ a ZS-CoT prompt \cite{kojima2022large} as our second baseline, instructing the model to provide a self-created step-by-step reasoning sequence with intermediate computation steps, instead of directly mapping high-level NLR to low-level HTML/CSS, which is effective for many tasks.

\subsection{Prompt Decomposition for GUI Generation (PDGG)}

With reference to the notion of enabling the model to perform step-by-step reasoning for generating GUIs instead of directly outputting code, we extend this idea by decomposing the instruction into multiple separated tasks and prompts \cite{khot2022decomposed} for GUI generation, which also follows a human expert approach more closely, as illustrated in Fig. \ref{fig:overview}B. Usually, requirements elicitation and gathering takes place initially and the human experts utilizes knowledge and previous experience to gain deeper insights into the requirements. Therefore, in the first step, we employ a feature extraction prompt, asking the LLM to provide a more detailed collection of GUI features with textual descriptions based on the high-level NLR. This reasoning step to perform initial feature analysis is important to form a deeper understanding of the different aspects of the GUI. Second, we instruct the LLM to derive implementation and design ideas for each of the features i.e. which GUI components and libraries to employ for implementing the feature. Third, we instruct the LLM to provide an overall layout structure and page design given the previously generated feature collection. With each step of the decomposed prompt pipeline, the initial more implicit NLR is extended by more and more information to form a more explicit representation of the GUI generation problem. Subsequently, this representation forms the input to the final LLM which is tasked with the translation of the entire text specification of the GUI into HTML/CSS. In addition to decomposing to task into sub tasks to provide more computation capabilities for each step, we also created a second variant of the described process by combining all instructions into a single prompt. This represents a custom ZS-CoT prompt of predefined domain-specific reasoning steps. 

\subsection{Retrieval-Augmented GUI Generation (RAGG)}
\label{sec:ragg}

The notion behind Retrieval-Augmented Generation (RAG) lies in integrating LLMs with their text generation and reasoning capabilities with the advantage of retrieval approaches - namely, providing rapid access to vast amounts of potentially relevant documents for effectively completing a generation task, which achieves SOTA results for many NLP problems \cite{weston2018retrieve, lewis2020retrieval, li2022survey}. Usually, the retrieved documents are provided as part of the prompting context in a ZS setting, enabling the LLM to complete tasks requiring knowledge which has not implicitly been stored in the LLM during pretraining. We adopt this idea and propose Retrieval-Augmented GUI Generation (RAGG), by leveraging the vast GUI prototyping knowledge embodied in a large-scale GUI repository for LLM-based GUI generation. Subsequently, we briefly delineate the employed GUI repository and its filtering, a novel LLM-based GUI re-ranking and filtering approach, to ensure high relevance of retrieved GUIs and our prompting approaches, to integrate knowledge embodied in the GUI screens for GUI generation. An overview of the RAGG approach is provided in Fig. \ref{fig:overview}C.

\subsubsection{GUI Repository}
\label{subsub:guirepo}

As described earlier, we closely follow the approach in \cite{kolthoff2023data}, exploiting \textit{Rico} \cite{deka2017rico} as the GUI repository for retrieval. \textit{Rico} is the largest publicly available GUI dataset for mobile apps, encompassing over 72k GUIs from 27 different diverse domains including GUI screenshots, hierarchy data and semantic annotations. \textit{S2W} \cite{wang2021screen2words} extends \textit{Rico} by providing over 112k brief high-level descriptions for over 22k \textit{Rico} GUIs. To improve the quality of the retrieved GUI data, we adopt their proposed filtering pipeline \cite{kolthoff2023data}, however, additionally filter GUIs with overlaying menus detected with a self-trained CNN classifier \cite{gu2018recent} (\textit{Precision}=.98 / \textit{Recall}=.70).

\subsubsection{LLM-Based GUI Re-Ranking}

To achieve NL-based GUI retrieval, we employ the \textit{SentenceBERT (S-BERT)} embedding model \cite{reimers2019sentence} to obtain low-dimensional dense embeddings for semantic retrieval, matching the NLR with the \textit{S2W} descriptions to ensure a small semantic gap between both representations \cite{kolthoff2023data, yuan2024maxprototyper}. To compute a ranking score over the entire dataset, we compute the cosine similarity between both embeddings i.e. $\textbf{Score}(\text{NLR, GUI}) = \mathbf{\cos}(\textbf{S-BERT}(\text{NLR}), \textbf{S-BERT}(\text{GUI$_{S2W}$}))$. Although this approaches enables rapid retrieval over the large-scale GUI repository, the relevance of retrieved documents is moderate, requiring a slower but more sophisticated re-ranking approach. While current specifically trained \textit{BERT-LTR} models significantly outperform baselines \cite{kolthoff2023data}, the achieved performance still indicates a large gap between human-level and model-based GUI relevance ranking. Since providing numerous irrelevant GUIs to the LLM for the GUI generation task unnecessarily consumes part of the context window and might confuse the LLM, we propose a novel multi-modal LLM-based GUI re-ranking approach, leveraging its reasoning and image understanding capabilities. In addition, recent research has shown that LLMs are capable of understanding the fine-grained semantics of GUIs for user story implementation detection and matching \cite{kolthoff2024interlinking}. Therefore, we exploit these capabilities for the re-ranking of GUI retrieval results. As shown in Fig. \ref{fig:overview}B, we first retrieve the top-\textit{n} relevant GUIs with \textit{S-BERT}, then apply a binary relevance filter using ZS prompting. Particularly, we encode each top-\textit{n} GUI screenshot and provide it to the LLM as part of the prompt, then instruct the LLM to decide whether the GUI is relevant to the given NLR. To ensure obtaining a high precision (and therefore avoiding inputting FPs later), we instructed the model to be critical and rather vote for non-relevant in case of ambiguity. Since the \textit{S-BERT} rank of GUIs not necessarily corresponds with fine-grained relevance, we additionally run a second-level ZS prompt instructing the LLM to provide a more fine-grained relevance score (on a scale from 1 (non-relevant) to 10 (most relevant)) and sort the binary filtered GUIs accordingly. To avoid inconsistencies in the relevance scores, we sampled the LLM multiple times and computed the average and standard deviation, which indicated consistent ratings. These filtered and sorted GUIs represent the basis for the GUI generation. We focused primarily on an LLM-based filtering mechanism, however, we additionally computed an entire re-ranking of the top-\textit{n} GUIs utilizing ZS prompting with all top-\textit{n} GUIs encoded in the context enabling a direct comparison to prior SOTA models on the NL-based re-ranking gold standard \cite{kolthoff2023data}.

\subsubsection{Prompting Approaches}

After obtaining the relevant GUIs according to the previously described procedure, we propose two distinct approaches for utilizing the GUIs for LLM-based GUI generation in combination with the NLR. First, \textit{(1)} we employ a ZS prompt for each GUI by encoding the screenshot in the prompt context of a multi-modal LLM and instruct the model to extract a feature collection from the GUI. Afterwards, we employ a second ZS prompt instructing the model to summarize and aggregate the \textit{k} feature collections of the top-\textit{k} GUIs. Semantically similar features should be aggregated and based on their frequency ranked higher in the collection. In addition, we employ a similar pipeline to extract design and layouts for each of the top-\textit{k} GUIs and subsequently aggregate them. Finally, aggregated feature and design collections are then provided in a ZS prompt instructing the model to generate HTML/CSS using the provided collections. Second, \textit{(2)} we directly encode the top-\textit{k} GUI screenshots in the context of a ZS prompt instructing the LLM to utilize the provided GUIs as inspirations for features, design and layout to generate the HTML/CSS for the respective GUI prototype.

\subsection{Self-Critique Looping for GUI Generation (SCGG)}

Self-Critique aims at enhancing the effectiveness of task solving of LLMs by employing the LLM itself to provide feedback to an LLM generated solution (i.e. criticize one's own prior output) \cite{saunders2022self}. Particularly, the effectiveness improves when providing critiques compared to directly ZS tasking the LLM to refine the response. For enhancing ZS-based GUI generation, we adapt the concept and propose Self-Critique Looping for GUI Generation (SCGG), an overview of the approach is illustrated in Fig. \ref{fig:overview}D. Our SCGG approach consists of two main components. First, \textit{(1)} a feedback LLM (i.e. GUI prototype critique) that utilizes the NLR and GUI prototype \textbf{P$_{i}$} to generate a feedback (i.e. critique) \textbf{F$_{i}$}. This feedback LLM is based on a ZS prompt, instructing the model to provide feedback on additional relevant GUI features that the current prototype neglects, improvement for the implementation of the current features and improvement for the design and layout of the overall prototype. In addition, we instruct the LLM to solely focus on textual descriptions and avoid providing code. Second, \textit{(2)} a GUI prototyping LLM that utilizes the NLR, prior GUI prototype \textbf{P$_{i}$} and the respective critique \textbf{F$_{i}$} to generate the subsequent GUI prototype \textbf{P$_{i+1}$}. In the ZS prompt, we instruct the model to improve the current GUI prototype by integrating the provided feedback and respond with a revised HTML/CSS without explanation. To summarize, the SCGG approach employs two components:
\begin{align}
  \text{Feedback-LLM:} \quad \mathbf{F}_i &= \text{ZS-LLM}(\mathbf{P}_i, \text{NLR}) \\
  \text{Prototyper-LLM:} \quad \mathbf{P}_{i+1} &= \text{ZS-LLM}(\mathbf{P}_i, \text{NLR}, \mathbf{F}_i)
\end{align}

\noindent This approach requires an initial GUI prototype in HTML/CSS to be able to provide the initial critique. Therefore, we provide the ZS instruction prototype as a starting point to further refine the prior generated GUI prototypes over \textit{k}-many iterations.

\begin{figure}
  \centering
  \includegraphics[width=0.44\textwidth]{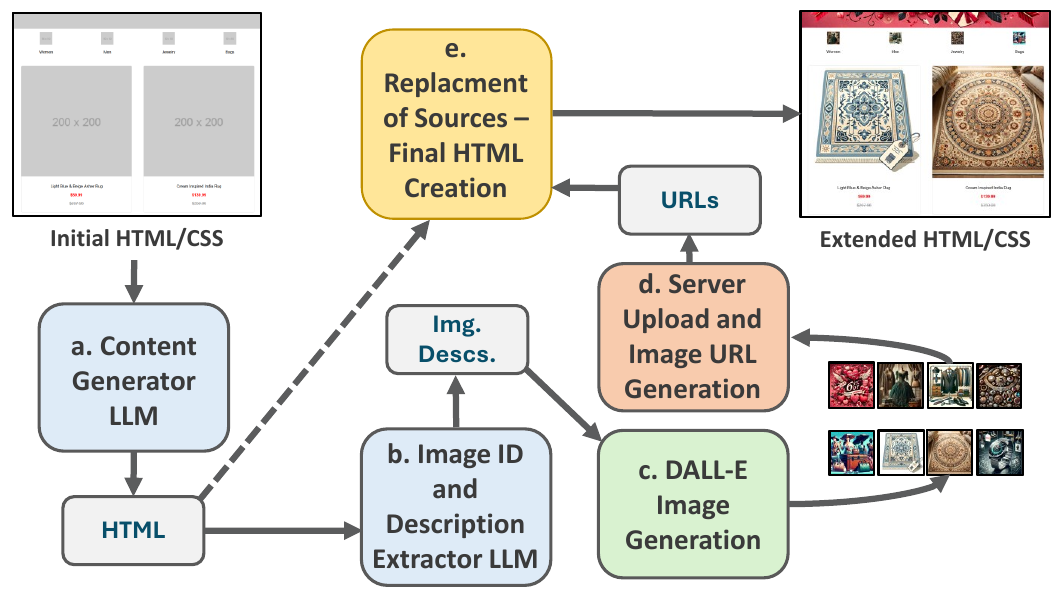}
  \caption{Overview of the LLM-based content generation}
	\label{fig:content_generation}
\end{figure}

\subsection{LLM-based GUI Content Generation}

To further improve the high-fidelity GUI prototypes, we propose a novel LLM-based GUI content generation pipeline illustrated in Fig. \ref{fig:content_generation}. Often, generated GUIs from the proposed ZS approaches neglect to include realistic data (or enough examples at all) and contain generic content. To enhance the realism and the appearance of the GUI prototypes using content, we first provide the base HTML/CSS to an LLM, instruct it to extend the prototype with realistic data and add \textit{id} attributes to each \textit{$<$img$>$} tag (e.g., add list items, related products etc.). Subsequently, we instruct another LLM to extract a collection of all images contained in the HTML and provide more detailed descriptions. We utilize an LLM for this step instead of directly extracting the \textit{$<$img$>$} tags with a library to allow the LLM to exploit the context to create better descriptions. These descriptions are then employed to generate images with DALL-E-3 \cite{betker2023improving}. Finally, the generated images are uploaded to a server and the generated URLs are incorporated into the HTML/CSS by matching previously extracted \textit{ids}.
\section{Evaluation}

This section presents the design and methodology of our experimental evaluation. In particular, the main goals of our evaluation are to \textit{(i)} measure the effectiveness of LLM-based GUI re-ranking and filtering and \textit{(ii)} assess the effectiveness of the different ZS prompting approaches for GUI generation. To this end, we state the following five research questions:

\begin{itemize}[leftmargin=0.2in]
    \item[] \textbf{RQ1}: How effective are LLM-based re-ranking and filtering techniques for NL-based GUI retrieval results?
    \item[] \textbf{RQ2}: Which LLM prompting approach is most effective for generating GUIs from short text descriptions?
        \item[] \textbf{RQ3}: How does the example number in the retrieval-augmented GUI generation impact its effectiveness?
    \item[] \textbf{RQ4}: How does the number of loops in the self-critique for GUI generation impact its effectiveness?
    \item[] \textbf{RQ5}: How does LLM-based content generation for GUIs influence the overall GUI effectiveness?
\end{itemize}

\subsection{GUI Descriptions Dataset}

In this section, we present the conducted procedure to collect our GUI description dataset to evaluate the posed RQs. In the first step \textit{(1)}, \textit{Rico} GUIs were sampled, for which textual descriptions were generated in the following step \textit{(2)} and then evaluated by crowd-workers to increase data quality \textit{(3)}.

\subsubsection{Rico GUI Sampling}

A GUI prototype selection was required before participants could create descriptions for them. We used the established \textit{Rico} GUI dataset \cite{deka2017rico}. To build the collection of \textit{Rico} GUIs serving as a foundation for the following data collection, we performed multiple steps to increase the quality and diversity of GUIs. First, we applied the same basic GUI filtering as described in Section \ref{subsub:guirepo}. To ensure the inclusion of feature-wise more comprehensive GUIs (to increase the difficulty for the GUI generation), we removed GUIs with less then seven different GUI component types. Then, three research assistants rated each GUI for six predefined categories 
(\textit{Clear Funct.}, 
\textit{English}, 
\textit{Safe for Work},
\textit{Errors}, 
\textit{Religious Content}, 
\textit{Simplicity}), generating 1050 ratings in total. On average, each GUI received 3 set of ratings. 
Consequently, 146 GUIs were admitted to our study.

\subsubsection{Collecting GUI Prototype Descriptions}

We recruited 30 participants (21 male, nine female), with an average age of 23.67 years ($\sigma$ = 3.25), from a university panel to create descriptions of GUI prototypes that later served as the basis for our automated GUI prototype generation.
We chose a lab-based setting instead of sourcing crowd-workers for this step to ensure that participants were not using LLMs to generate the GUI prototype descriptions. Recent work \cite{veselovsky2023artificialintelligence} has shown that crowd-workers broadly use LLMs for text-generation tasks. Each participant created 20 GUI prototype descriptions. 600 GUI prototype descriptions were created in total, with an average of 4.11 descriptions for each of the 146 individual GUIs.
We then had a research assistant examine the created GUI prototype descriptions to identify descriptions we would have to sort out (e.g., because of sensitive content or personal data). Consequently, 26 descriptions were sorted out. Our final collection amounted to 574 GUI prototype descriptions.

\subsubsection{Evaluating GUI Prototype Descriptions}
Next, we asked crowd-workers to rate the created GUI descriptions, filtering for quality and ensuring understandability. Additionally, we expected the GUI descriptions to not solely to represent a listing of single features since such a granular feature level is usually unavailable when generating first GUIs. We also required descriptions that allow understanding of not only the individual GUI page (essential functionality), but also the overall app.
To rate the previous GUI descriptions, we invited 30 crowd-workers from \textit{Prolific} \cite{palan2018prolific}, requiring English skills (B2) and experience in UI/UX design. We excluded two participants for failing our attention checks, leaving us with 28 crowd-workers (20 male, seven female, and one non-binary). Participants had an average age of 30.50 ($\sigma$ = 7.13), self-reported 3.79 ($\sigma$ = 3.55) years of experience creating and 2.71 years ($\sigma$ = 3.08) of experience evaluating visual design.

Crowd-workers rated 574 GUI descriptions for 146 GUIs (on average, 3.93 descriptions for one GUI). Participants, on average, created 1.95 ratings per GUI description on 9-point Likert scales (e.g., \textit{Description Understandability}, \textit{App Purpose Understandability}, full list in our repository \cite{zs-prompting-github}).
To select a single suitable GUI description per GUI, we created a weighted average of ratings for each created GUI description. For each GUI, we then selected the description which was rated the highest. From the 146 high-quality GUI descriptions, we randomly sampled ten examples as our validation dataset, which we used to conduct internal experiments. Subsequently, we randomly sampled 50 descriptions from the remaining examples to obtain our final test dataset. For feasibility of conducting the expensive and labour-intensive annotation of generated GUIs by crowd-workers, we restricted the test set to 50 examples, while still encompassing diverse descriptions.

\subsection{RQ1: Effectiveness of LLM-based GUI Re-Ranking}

\subsubsection{Datasets}

We evaluated our GUI re-ranking models using a publicly available gold standard for evaluating NL-based GUI ranking \cite{kolthoff2023data}, enabling a direct comparison to SOTA methods. In addition, we utilized our test set from the previous section of 50 GUI descriptions and conducted top-20 retrieval with the \textit{S-BERT} model. To avoid bias, we discarded the GUIs from the \textit{Rico} GUI retrieval dataset which we employed to collect the descriptions. To evaluate the binary LLM-based relevance filtering, we annotated the retrieved GUIs with three annotators for binary relevance and computed the majority vote as the ground truth (Fleiss Kappa = .403).

\subsubsection{Evaluation Metrics}

On the gold standard, we computed standard information retrieval metrics including Average Precision (AP), Mean Reciprocal Rank (MRR), Precision@\textit{k}, HITS@\textit{k} and NDCG@\textit{k} (details in \cite{kolthoff2023data}). On the second dataset, we use Precision (P), Recall (R) and F1-Measure (F1).

\subsubsection{Model Setup}

We used the most recent \textit{GPT-4o} model  (128k tokens context length, accessed in July, 2024), which is a multi-modal extension of the previous \textit{GPT-4} model \cite{openai2023gpt4}. This model achieves SOTA performance on many text evaluation and visual perception tasks. We evaluated three different temperature settings (t: .25$|$.50$|$.75) on the gold standard.

\subsection{RQ2: Effectiveness of ZS Prompting for GUI Generation}
\label{sec:rq2}


\subsubsection{Annotation of Generated GUI Prototypes}

The evaluation of the approaches focuses on assessing the created GUIs. We utilized working prototypes in HTML/CSS for this assessment, and humans evaluated the GUIs based on our metrics and the GUI description. For this purpose, we hired crowd-workers with UI/UX experience on the crowd-working platform \textit{Prolific} \cite{prolific_academic_ltd_prolific_nodate}. Participants on \textit{Prolific} offer better quality compared to other crowd-work platforms \cite{douglas2023data}. We were also able to require a high approval rate ($>99\%$), minimum number of previously completed tasks ($>30$), experience in UI/UX design (self-reported), and English skills.

\definecolor{lightgray}{rgb}{0.93, 0.93, 0.93}

\begin{table}[t]
\footnotesize
\caption{Likert scale items and subjective measures.}
\label{tab:questions_liker}
\begin{tabular}{|l|l|l|}
\hline
\multicolumn{1}{|c|}{\textbf{}} & \multicolumn{1}{c|}{\textbf{Likert Scale  Question}} & \multicolumn{1}{c|}{\textbf{Sub. Measure}} \\ \hline \hline

\rowcolor{lightgray}
\textbf{A} & \begin{tabular}[c]{@{}l@{}}This GUI prototype fulfills all functions \\ defined in the GUI task description.\end{tabular} & Feature Compl. \\ 

\rowcolor{white}
\textbf{B} & \begin{tabular}[c]{@{}l@{}}The GUI prototype's feature set is extensive, \\ clearly exceeding the GUI task description \\ by implementing additional useful functions.\end{tabular} & Feature Width \\

\rowcolor{lightgray}
\textbf{C} & \begin{tabular}[c]{@{}l@{}}The features were perfectly implemented for \\ the given GUI prototyping task.\end{tabular} & Feature Impl. \\ 

\rowcolor{white}
\textbf{D} & \begin{tabular}[c]{@{}l@{}}The organization of information on the GUI \\ prototype page is clear.\end{tabular} & \begin{tabular}[c]{@{}l@{}}Information \\ Organization\end{tabular} \\ 

\rowcolor{lightgray}
\textbf{E} & The visual design of the GUI prot. is appealing. & Visual Appeal \\ 

\rowcolor{white}
\textbf{F} & This GUI prototype has only minimal errors. & Errors in GUI \\ 

\rowcolor{lightgray}
\textbf{G} & Overall, I am satisfied with this GUI prototype. & Overall Satisf. \\ 

\rowcolor{white}
\textbf{H} & \begin{tabular}[c]{@{}l@{}} The presented GUI prototype looks like \\ a screen from a complete app.\end{tabular} & \begin{tabular}[c]{@{}l@{}}Complete App \\ \end{tabular} \\ \hline
\end{tabular}
\end{table}

Out of 126 survey submissions, we meticulously screened out 25 participants, ensuring a high-quality dataset (eight participants did not complete the survey, and 17 participants failed one or more attention checks). The following data is reported for the 101 participants (73 male, 27 female, and one non-binary) accepted to our study. Participants had an average age of 30.86 years ($\sigma$ = 9.21), 4.60 years of experience creating a visual design (self-expressed, $\sigma$ = 4.61), and 3.58 years of experience evaluating visual design (self-expressed, $\sigma$ = 3.72). 
49\% of our participants reported \textit{somewhat high} to \textit{very high} (85\% \textit{medium} or higher) knowledge of creating GUI prototypes and 39\% \textit{somewhat high} to \textit{very high} (78\% \textit{medium} or higher) skills in creating GUI prototypes. Participants received USD 15.25 for participating in our 65 minute study.

\begin{table*}[!t]
\footnotesize
\caption[Evaluation Results]{Gold standard results of previous SOTA GUI re-ranking approaches in comparison to LLM-based GUI re-ranking}
\centering
\setlength\tabcolsep{6pt} 
\begin{tabular}{l|c|c|cccc|cccc|ccccc}
\toprule
 & \multicolumn{1}{c|}{\textbf{AR}} & \multicolumn{1}{c|}{\textbf{MRR}} & \multicolumn{4}{c|}{\textbf{Precision@k}}  & \multicolumn{4}{c|}{\textbf{HITS@k}} & \multicolumn{4}{c}{\textbf{NDCG@k}}\\
 \cmidrule{2-2} \cmidrule{3-3} \cmidrule{4-7} \cmidrule{8-11} \cmidrule{11-15}
 & AP & MRR& $P@3$ & $P@5$ & $P@7$ & $P@10$ & $H@1$ & $H@3$ & $H@5$ & $H@10$ & $N@3$ & $N@5$ & $N@10$ & $N@15$ \\ \midrule
 BERT-LTR (1) & .486 &  .618 &       .377 &  .350 &   .307 &  .269 &   .460 &     .710 &     .860 &     .980 &        .530 &     .560 &     .634 &     .697 \\
 \rowcolor{lightgray} BERT-LTR (2) &  .501 &  .631 &       .400 &    .340 &   .304 &  .281 &  .440 &     .750 &    .910 &     .980 &     .543 &    .556 &    .636 &     .701 \\
 BERT-LTR (3) &.499 &  .626 &       .363 &  .354 &  .317 &  .287 &  .450 &     .730 &     .860 &     1.00 &     .517 &    .554 &  .646 &     .694 \\ \midrule

 \rowcolor{lightgray} LLM-Rerank$_{t=0.25}$ & .805 &  .893 &       \textbf{.727} &  \textbf{.578} &   \textbf{.463} &  \textbf{.347} &   .820 &     .970 &     .990 &     \textbf{1.00} &     .845 &     \textbf{.840} &     .859 &     \textbf{.894} \\
 LLM-Rerank$_{t=0.50}$ &  .801 &  .898 &       \textbf{.727} &    .556 &   .444 &  .343 &  .830 &     .960 &    .980 &     \textbf{1.00} &        .852 &    .831 &    .857 &     .886 \\
 \rowcolor{lightgray}LLM-Rerank$_{t=0.75}$ &\textbf{.818} &  \textbf{.903} &       .723 &  .560 &  .446 &  .343 &  \textbf{.840} &     \textbf{.980} &     \textbf{1.00} &     \textbf{1.00} &      \textbf{.855} &    .831 &  \textbf{.862} &     .890 \\
 \midrule
\bottomrule
\end{tabular}
\label{tab:results_rq1}
\end{table*}

We conducted a comprehensive evaluation of the approaches and baseline models by creating interactive GUIs for each approach and baseline and each of 50 sampled GUI descriptions. Each GUI was then annotated three times, resulting in 3,000 annotations.
In our survey-based evaluation, participants first received a detailed explanation of what to evaluate and then answered comprehension checks. Participants were shown GUI descriptions, eight Likert scale items, and a link to the interactive GUI prototype. We ensured that no participant received multiple GUIs generated from the same GUI description so that similar GUIs would not bias participants. 

\subsubsection{Evaluation Metrics}
To measure effectiveness of generated GUI prototypes, we decided on evaluating subjective satisfaction. Table \ref{tab:questions_liker} presents the measured Likert items (9-point scale from \textit{Strongly Disagree} to \textit{Strongly Agree}).

\subsubsection{Model Setup}

To conduct the experiments, we employed the identical \textit{GPT-4o} configuration for the LLM as described in RQ1, however, set a consistent temperature of $t = .50$ for all models to balance creativity and consistency in the generated outputs. We then compared four groups of models on significant differences, namely \textit{(i)} baselines, \textit{(ii)} prompt decomposition, \textit{(iii)} RAGG and \textit{(iv)} SCGG. We also tested for differences between \textit{(i)} the decomposed prompt and the engineered ZS-CoT prompt, \textit{(ii)} RAGG$_{k=3}$ with direct encoding and RAGG$_{k=3}$ with explicit feature and layout extraction.

\subsection{RQ3: Effectiveness of Example Number on RAGG}

To evaluate the influence of the number \textit{k} of retrieved GUIs for RAGG, we employed the approach with direct encoding of top-\textit{k} GUIs and the identical \textit{GPT-4o} configuration (cp. RQ2). We therefore used this model with four different example numbers that are included in the context ($k = 1,3,5,7$). We restricted the maximum number tested in this setup, since only a smaller fraction of the entire dataset would be influenced by it (only 30\% of the top-20 retrieval sets of the GUI descriptions contain seven or more relevant GUI screens). Additionally, we computed the results on a subset of the test set, to ensure that each of the included examples has at least seven relevant GUIs.

\subsection{RQ4: Effectiveness of Number of Loops in SCGG}

To evaluate the number of loops in the SCGG approach for the effectiveness of GUI generation, we employed the ZS instruction output as the input and the identical \textit{GPT-4o} configuration (cp. RQ2). Then, we evaluated this model with four different settings ($k = 1,2,3,4$) as the number of loops.

\subsection{RQ5: Effectiveness of GUI Content Generation}

To evaluate the influence of the GUI content generation approach on the effectiveness of generated GUIs, we used the identical \textit{GPT-4o} configuration (cp. RQ2) and employed two models per group with and without content generation, respectively. Therefore, we employed \textit{(i)} the ZS-CoT for the baselines, \textit{(ii)} the decomposed prompting approach for PDGG, \textit{(iii)} RAGG$_{k=3}$ with direct encoding and \textit{(iv)} SCGG$_{k=3}$ with ZS instruction prototyping output as the initial GUI prototype.

\section{Results \& Discussion}

In this section, we present the results of the previously posed research questions and provide a more detailed discussion.

\subsection{RQ1: Effectiveness of LLM-Based GUI Re-ranking}

Table \ref{tab:results_rq1} shows the results of the LLM-based GUI re-ranking models with varying temperature on the gold standard. As can be observed, our proposed method significantly outperforms the previous BERT-LTR SOTA GUI re-ranking models across all metrics. In particular, an average precision of .818 (compared to .501 from the best BERT-LTR model) of our third model indicates that our approach is able to produce only a small number of FPs. This substantial improvement across all metrics indicates the strength of the multi-modal LLM-based ZS prompting approach for GUI re-ranking. This adds on to the findings of previous research \cite{kolthoff2024interlinking} that LLMs can perform verification of user story implementation in GUI prototypes requiring a detailed understanding of the functionality and semantics of the GUIs. In addition, we computed the LLM-based binary relevance judgement results on the actual test dataset and obtained $P = .757$, $R = .814$ and $F1 = .784$, which supports the previous findings in the high annotation quality of the LLM-based method. To gain deeper insights into the FPs predicted by the model, we randomly sampled 15 FPs and analyzed them, a detailed overview of the examples including model explanations and comments is available at our repository \cite{zs-prompting-github}. This analysis shows that the FPs frequently are still quite relevant, often solely minor features were missing or different, which led the model to produce false predictions.

\begin{mybox}
\textbf{Answer to RQ1:} LLM-based GUI re-ranking significantly outperforms SOTA GUI re-ranking models
\end{mybox}

\subsection{RQ2: Effectiveness of ZS Prompting for GUI Generation}

\begin{figure*}
 \includegraphics[width=\textwidth]{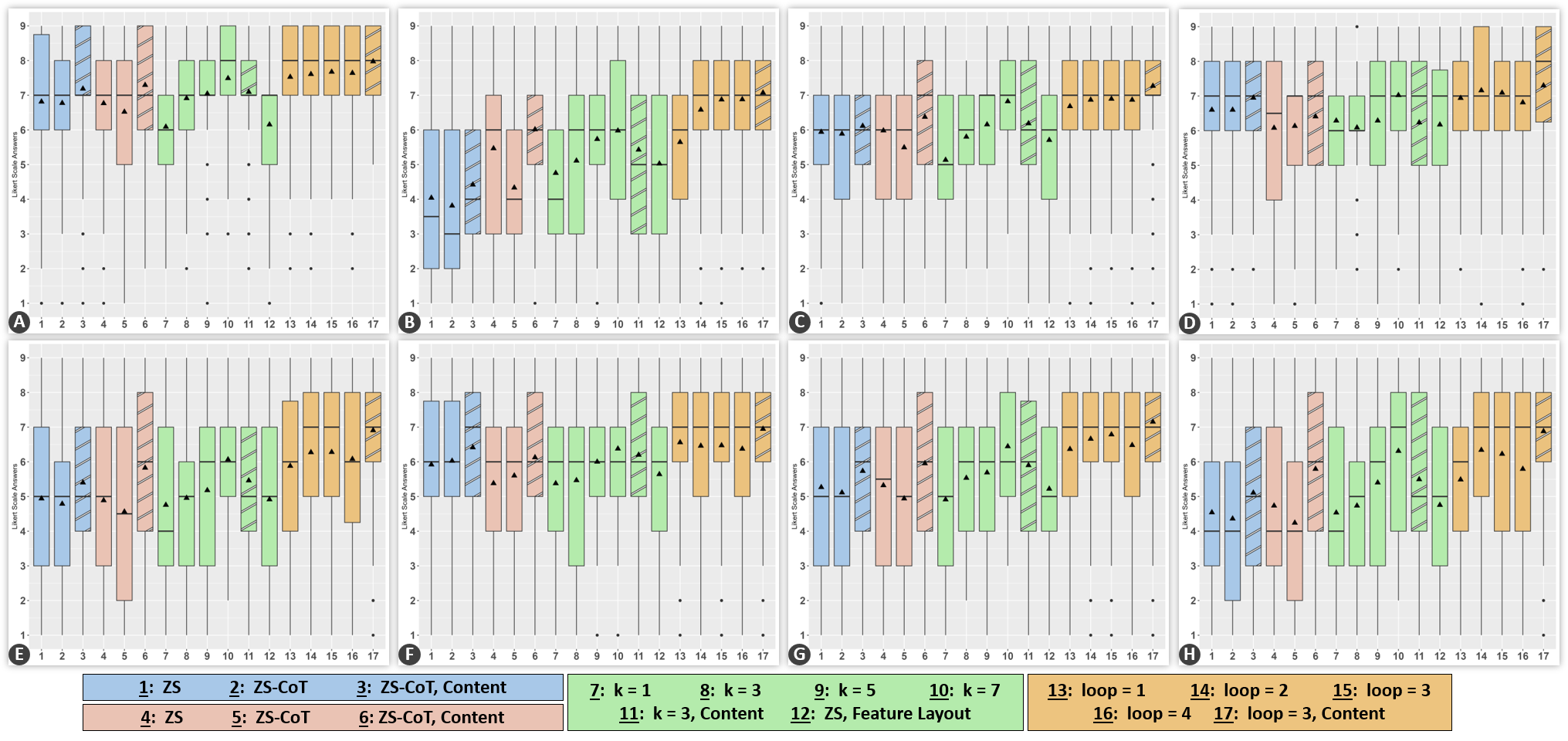}
  \caption{Crowd-workers' GUI ratings, generated  for baseline (1-3, blue), PDGG (4-6, red), RAGG (7-12, green), SCGG (13-17, orange). A: \textit{Feature Compl.}, B: \textit{Feature Width}, C: \textit{Feature Impl.}, D: \textit{Inform. Organiz.}, E: \textit{Visual Appeal}, F: \textit{Errors in GUI}, G: \textit{Overall Satisfaction}, H: \textit{Complete App}. Hatched bars represent models with GUI content generation, triangles represent means.}
	\label{fig:results_main}
\end{figure*}

To assess the effectiveness of our approaches for GUI generation, we evaluated the subjective ratings of our crowd-workers. Fig. \ref{fig:results_main} presents boxplots for the ratings. 

Subsequently, we briefly summarize the results for the baseline (ZS, ZS-CoT), PDGG, RAGG (k = 1-7) and SCGG (loop number 1-4) approaches, not considering GUI content generation.

SCGG significantly outperforms our baseline and PDGG models when evaluating which approach best fulfills all functions defined in the GUI prototype descriptions. 
Our SCGG approaches were all rated significantly higher than the baseline approaches for producing a feature set that exceeds the GUI task description with additional useful components. For this capability, RAGG approaches utilizing more than five examples were also rated significantly higher than the baseline models. PDGG (in the first configuration, sequentially) also significantly performed better than our baseline. 
SCGG approaches were rated significantly higher for perfect feature implementation than baseline, and both PDGG approaches (except SCGG with a loop number greater than one). 
For loop numbers two and three, all of our SCGG models were rated significantly higher for information organization in direct contrast to both PDGG approaches. 
Participants significantly rated GUIs generated with SCGG as visually more appealing in direct contrast to baseline and PDGG models. Similarly, SCGG performed significantly better than GUIs generated with RAGG, utilizing three or fewer examples. 
Participants rated GUIs generated with SCGG as significantly less error-prone than GUIs generated with our PDGG models.
The participants' overall satisfaction with our GUIs was significantly higher for SCGG in direct contrast to all baseline and PDGG approaches. SCGG also outperformed RAGG for overall satisfaction when only one example was utilized. When seven examples were utilized, RAGG was rated significantly better than the baseline models for overall satisfaction.
Participants significantly rated GUIs as looking like a screen from a complete app when SCGG (more than two loops) or RAGG with seven utilized examples rather than the baseline or PDGG approaches were used to create the GUIs. 
PDGG in separate prompts led to significantly better results for feature extensiveness and feature implementation compared to a single prompt (p-values of 0.00001 and 0.02518). We did not find significant effects for feature extraction for RAGG approaches.

\begin{figure*}
  \centering
 \includegraphics[width=0.97\textwidth]{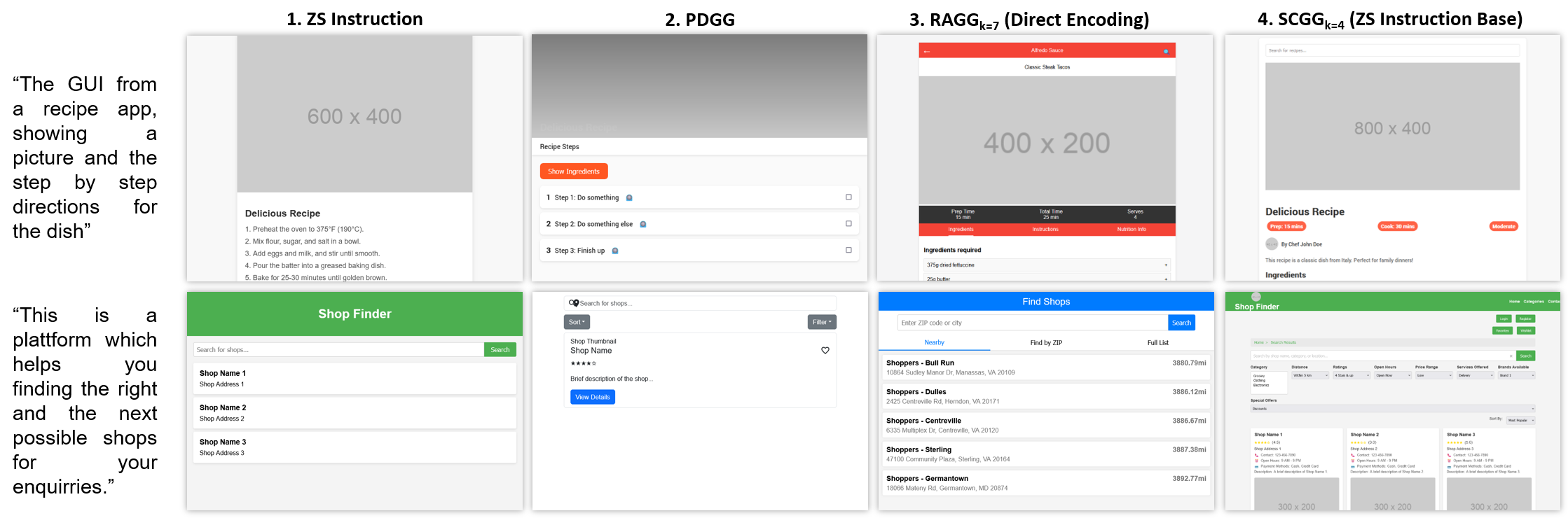}
  \caption{Generated GUI prototypes for \textit{(1)} ZS instruction, \textit{(2)} Prompt Decomposition, \textit{(3)} RAGG$_{k=7}$ and \textit{(4)} SCGG$_{k=4}$}
	\label{fig:examples}
\end{figure*}

To better understand the defects of LLM-generated GUI prototypes, we manually investigated a sample of GUIs with the lowest \textit{Errors in GUI} (F) score. The identified defects can be structured into potentially non-exhaustive three categories. First, \textit{(i)} defects related to the \textit{layout} occurred frequently. This includes misalignment of components (e.g., between labels and their respective check boxes), component overlapping (e.g., labels and radio buttons or icons with search bar), overflow of content (e.g., images and text that exceed their grouping components) and underflow of content (e.g., list items that only span half of the container width). Second, \textit{(ii)} defects related to \textit{function} including misinterpretation of requirements (e.g., navigation app provides buttons to navigate through the app), missing features (e.g., upcoming shows not represented in a calendar, no header or footer), non-usual implementation of features (e.g., implementing a save or share functionality with text buttons instead of icons) and broken features (e.g., broken image or icon source links). Third, \textit{(iii)} defects related to inconsistent styling of the GUI (e.g., inconsistent color schema or layouts). Fig. \ref{fig:examples} shows two descriptions taken from our test set and generated GUIs for models from each group. The examples illustrate a clear improvement regarding the amount of features from the baseline to the SCGG approach.

\begin{mybox}
\textbf{Answer to RQ2:} SCGG significantly outperforms ZS, ZS-CoT and PDGG across most evaluated aspects.
RAGG approaches with more than five utilized examples were significantly higher rated for ZS, ZS-CoT for \textit{Feature Width} and presenting a \textit{Complete App}
\end{mybox}

\subsection{RQ3: Effectiveness of Example Number on RAGG}

Table \ref{tab:results_rag_k} shows the test results for the RAGG approach with varying values for \textit{k}. There is a significant improvement for feature-related items (A)-(C) between \textit{k}=1 and \textit{k}=5, but no significant improvement over other items indicating that providing more relevant examples helps the LLM to implement better and more features in the generated GUIs. Comparing \textit{k}=1 and \textit{k}=7, we can observe a significant improvement over all items. This indicates that with even more GUIs, the LLM can exploit the relevant GUIs not only for features but also for improving the overall GUI design. A similar improvement can be observed between \textit{k}=3 and \textit{k}=7 except for the \textit{Feature Width} (B), which indicates that the additional GUIs mainly act as design inspirations for the LLM. Between neighbouring steps, no significance could be observed (except 1-3 item (A)).

\begin{mybox}
\textbf{Answer to RQ3:} A large example number for RAGG ({\textit{k}=7}) significantly outperforms small example numbers (\textit{k}=1, 3) for the effectiveness of GUI generation
\end{mybox}

\begin{table}[t]
\caption{p-Values (Wilcoxon Signed-Rank Test) for different RAGG \textit{k}-values and eight evaluated aspects (cleaned RAGG set, N=45). A: \textit{Feature Compl.}, B: \textit{Feature Width}, C: \textit{Feature Impl.}, D: \textit{Inform. Organization}, E: \textit{Visual Appeal}, F: \textit{Errors in GUI}, G: \textit{Overall Satisfaction}, H: \textit{Complete App}.}
\label{tab:results_rag_k}
\centering
\small
\begin{tabular}{|c|ll|ll|ll|}
\hline 
\rule{0pt}{2.2ex} 
\textbf{k}  & \multicolumn{2}{c|}{\textbf{3}} & \multicolumn{2}{c|}{\textbf{5}} & \multicolumn{2}{c|}{\textbf{7}} \\ \hline \hline
\textbf{\begin{tabular}[c]{@{}c@{}}\textbf{1}\end{tabular}} & \cellcolor[HTML]{F5F5F5}\begin{tabular}[c]{@{}l@{}}\scriptsize \textbf{.017} (A)\\ \scriptsize .306 (B)\\ \scriptsize .123 (C)\\ \scriptsize .607 (D)\end{tabular} & \begin{tabular}[c]{@{}l@{}}\scriptsize .699 (E)\\ \scriptsize .964 (F)\\ \scriptsize .204 (G)\\ \scriptsize .649 (H)\end{tabular} & \cellcolor[HTML]{F5F5F5}\begin{tabular}[c]{@{}l@{}}\scriptsize \textbf{.008} (A)\\ \scriptsize \textbf{.010} (B)\\ \scriptsize \textbf{.030} (C)\\ \scriptsize .910 (D)\end{tabular} & \begin{tabular}[c]{@{}l@{}}\scriptsize .333 (E)\\ \scriptsize .160 (F)\\ \scriptsize .061 (G)\\ \scriptsize .065 (H)\end{tabular} & \cellcolor[HTML]{F5F5F5}\begin{tabular}[c]{@{}l@{}}\scriptsize \textbf{.001} (A)\\ \scriptsize \textbf{.020} (B)\\ \scriptsize \textbf{.001} (C)\\ \scriptsize \textbf{.040} (D)\end{tabular} & \begin{tabular}[c]{@{}l@{}}\scriptsize \textbf{.002} (E)\\ \scriptsize \textbf{.018} (F)\\ \scriptsize \textbf{.001} (G)\\ \scriptsize \textbf{.000} (H)\end{tabular} \\ \hline \hline
\textbf{\begin{tabular}[c]{@{}c@{}}\textbf{3}\end{tabular}} & \cellcolor[HTML]{F5F5F5} &  & \cellcolor[HTML]{F5F5F5}\begin{tabular}[c]{@{}l@{}}\scriptsize .590 (A)\\ \scriptsize .153 (B)\\ \scriptsize .379 (C)\\ \scriptsize .739 (D)\end{tabular} & \begin{tabular}[c]{@{}l@{}}\scriptsize .660 (E)\\ \scriptsize .157 (F)\\ \scriptsize .670 (G)\\ \scriptsize .169 (H)\end{tabular} & \cellcolor[HTML]{F5F5F5}\begin{tabular}[c]{@{}l@{}}\scriptsize \textbf{.050} (A)\\ \scriptsize .080 (B)\\ \scriptsize \textbf{.004} (C)\\ \scriptsize \textbf{.004} (D)\end{tabular} & \begin{tabular}[c]{@{}l@{}}\scriptsize \textbf{.022} (E)\\ \scriptsize \textbf{.042} (F)\\ \scriptsize \textbf{.029} (G)\\ \scriptsize \textbf{.001} (H)\end{tabular} \\ \hline %
\end{tabular}
\end{table}

\subsection{RQ4: Effectiveness of Number of Loops in SCGG}

The conducted statistical tests between the SCGG approaches with increasing number of loops \textit{k} showed no significance among most items. However, comparing \textit{k=2,3,4} with \textit{k}=1, we could observe a significant improvement for \textit{Feature Width} (B) and \textit{Complete App} (H) (except for \textit{k}=4). This indicates that the SCGG model performs plenty of GUI prototype improvements from the initial prototype to the first iteration and afterwards mainly new features are added and the prototype is extended for more app completeness (e.g., adding header, navigation and footer). This could potentially be improved by dynamically focusing the feedback LLM more on currently underdeveloped parts of the GUI prototype.

\begin{mybox}
\textbf{Answer to RQ4:} Increasing the number of rounds in SCGG provides mainly improvements on feature-related and app completeness aspects
\end{mybox}

\subsection{RQ5: Effectiveness of GUI Content Generation}

Table \ref{ttab:results_content} shows the test results of the four different models with content generation each compared with the same model without content generation. As can be observed, \textit{Visual Appeal} (E), \textit{Complete App} (H), \textit{Overall Satisfaction} (G) (except SCGG) are significantly improved over the same models without content. This indicates that our automatic ZS prompt-based GUI content generation pipeline is effective to improve the appearance of the GUI prototypes. In addition, also the \textit{Feature Completeness} (A) (except ZS-CoT) is improved. One reason might be that e.g., only a single list item with generic content might not exactly implement the full requested functionality.

\begin{mybox}
\textbf{Answer to RQ5:} LLM-based GUI content generation significantly improves \textit{Visual Appeal}, \textit{Overall Satisfaction} and \textit{Complete App} aspects among most prompting approaches of created GUI prototypes
\end{mybox}
\section{Limitations}

\definecolor{lightgray}{rgb}{0.92, 0.92, 0.92}

\begin{table}[t]
\caption{Comparing ZS-CoT, PDGG, RAGG$_{k=3}$, SCGG$_{k=3}$ pairwise with the same approach and content generation (p-Values, Wilcoxon Signed-Rank Test, N=150).}
\label{ttab:results_content}
\begin{tabular}{|l||lllc|}
\hline
 & \multicolumn{1}{c|}{\begin{tabular}[c]{@{}c@{}} \rule{0pt}{2.5ex} ZS CoT\end{tabular}} & \multicolumn{1}{c|}{\begin{tabular}[c]{@{}c@{}} \rule{0pt}{2.5ex} PD ZS\end{tabular}} & \multicolumn{1}{c|}{\begin{tabular}[c]{@{}c@{}} \rule{0pt}{2.5ex} RAG ZS 3\end{tabular}} & \begin{tabular}[c]{@{}c@{}} \rule{0pt}{2.5ex} SC ZS 3\end{tabular} \\  
 \hline 
\rowcolor{white}
(A) Feature Compl. & \multicolumn{1}{c|}{0.061} & \multicolumn{1}{c|}{\textbf{0.008}} & \multicolumn{1}{c|}{\textbf{0.022}} & \textbf{0.020} \rule{0pt}{2.2ex} \\ 
\rowcolor{lightgray}
(B) Feature Width & \multicolumn{1}{c|}{\textbf{0.040}} & \multicolumn{1}{c|}{\textbf{0.020}} & \multicolumn{1}{c|}{\textbf{0.005}} & 0.215 \rule{0pt}{2.2ex} \\ 
\rowcolor{white}
(C) Feature Implem. & \multicolumn{1}{c|}{0.399} & \multicolumn{1}{c|}{0.072} & \multicolumn{1}{c|}{\textbf{0.014}} & \textbf{0.047} \rule{0pt}{2.2ex} \\ 
\rowcolor{lightgray}
(D) Inform. Orga. & \multicolumn{1}{c|}{0.077} & \multicolumn{1}{c|}{0.293} & \multicolumn{1}{c|}{0.503} & 0.337 \rule{0pt}{2.2ex} \\
\rowcolor{white}
(E) Appeal. Design & \multicolumn{1}{c|}{\textbf{0.019}} & \multicolumn{1}{c|}{\textbf{0.000}} & \multicolumn{1}{c|}{\textbf{0.007}} & \textbf{0.014} \rule{0pt}{2.2ex} \\ 
\rowcolor{lightgray}
(F) Minimal Errors & \multicolumn{1}{c|}{0.090} & \multicolumn{1}{c|}{\textbf{0.005}} & \multicolumn{1}{c|}{\textbf{0.013}} & \textbf{0.034} \rule{0pt}{2.2ex} \\ 
\rowcolor{white}
(G) Overall Satisf. & \multicolumn{1}{c|}{\textbf{0.013}} & \multicolumn{1}{c|}{\textbf{0.007}} & \multicolumn{1}{c|}{\textbf{0.003}} & 0.087 \rule{0pt}{2.2ex} \\ 
\rowcolor{lightgray}
(H) Complete App & \multicolumn{1}{c|}{\textbf{0.010}} & \multicolumn{1}{c|}{\textbf{0.000}} & \multicolumn{1}{c|}{\textbf{0.001}} & \textbf{0.013} \rule{0pt}{2.2ex} \\ \hline
\end{tabular}
\end{table}

While we presented several advantages and discussed the effectiveness of our proposed ZS prompting approaches for GUI generation, they also possess several limitations. First, the LLM-based GUI re-ranking approaches significantly outperform SOTA methods, however, still produce FPs in some cases. Although the FPs are usually not entirely irrelevant to the posed problem, these false instances might still negatively influence the GUI generation effectiveness of the RAGG approach. Therefore, RAGG is highly dependent on the retrieval method. For descriptions for which neither the retrieval model nor LLM-based filtering are able to find relevant GUIs or the GUI repository lacks relevant GUIs overall, the GUI generation falls back to a ZS instruction approach. In addition, RAGG would potentially work most effective if the knowledge embodied in the external GUI repository and internally saved in the LLM are complementary to each other. Due to the fact that GUI retrieval on \textit{Rico} as well as the LLM are both data-driven models, the distribution of seen GUI knowledge could be similar. However, in a more particular domain with custom characteristics of the GUI prototypes, which the LLM probably will less likely have stored, the RAGG approach might be able to outperform LLM-only methods. While SCGG shows the highest effectiveness for generating GUIs, it requires also to consumes a large amount of tokens while providing the critique and re-generating the prototype. This could potentially be enhanced in the future by a more sophisticated re-generation approach e.g., updating only effected segments of the GUI prototype. Finally, while our approach generates high-fidelity GUI prototypes where large parts are interactive, some components are non-interactive, which would require JavaScript (e.g., buttons, navigation items, links, sorting functionality etc.).
\section{Threats to Validity}

\subsubsection{Internal Validity}

One potential threat to internal validity is the relationship between measured items. In particular, a larger number of features in the generated prototypes might increase the difficulty of properly organizing the information, increase the number of feature implementation errors and increase the difficulty to create appealing visual designs. Since the simpler baseline methods usually only implement a small number of features and the other metrics are not normalized on feature count or complexity, this could potentially favor the simpler methods in the mentioned items. Another threat is the selection of participants to evaluate the generated GUI prototypes. To reduce bias, we solely included participants with self-reported experience in GUI design, included only participants that were fluent in English, having at least 30 prior submissions and an approval rate of over 99\%. In addition, recent research showed that crowd workers on Prolific create better data quality in comparison to other popular crowd-working platforms \cite{douglas2023data}. Furthermore, we excluded annotations from workers that failed one or more of overall five exposed attention checks.

\subsubsection{External Validity}

One threat to external validity is the description dataset that we employed in our experiments. Since the descriptions are based on the \textit{Rico} GUI dataset, the generalizability of our findings might be restricted. For example, GUIs from special domains which are rare might not be included in the \textit{Rico} dataset. For these rare GUIs, the descriptions we could obtain from users might also contain more ambiguity or obscurity. To ensure a representative sample of functional-wise more complex GUIs employed for creating the descriptions, we filtered the GUIs by containing at least seven unique GUI component types and then applied random sampling. Furthermore, we asked annotators to detect functionally simple GUIs (e.g., login), which we then excluded by majority voting. This increased the inclusion of more complex GUIs, increasing the difficulty of the GUI generation task. In addition, for more then 20\% of the descriptions the retrieval approach is not able to find relevant GUIs, showing the complexity of descriptions and the inclusion of rare GUIs.

\section{Related Work}

\subsubsection{GUI Generation}

Earlier research proposed several machine learning based GUI generation approaches. \textit{Variational transformer networks} \cite{arroyo2021variational} and \textit{LayoutTransformer} utilizing self-attention \cite{gupta2021layouttransformer} have been proposed to generate high-level layouts, including basic low-fidelity GUI layouts with abstract GUI component rectangles. In line with previous approaches, mixing \textit{transformer encoder-decoder} with \textit{Gaussian mixture models} \cite{bishop1994mixture} for generating abstract low-fidelity GUIs from text has been proposed before \cite{huang2021creating}. While these approaches can provide coarse design guidance, the generated layouts lack functional and design details compared to the interactive prototypes generated by our approach. Moreover, \textit{GUIGAN} \cite{zhao2021guigan} utilizes \textit{generative adversial networks (GANs)} \cite{goodfellow2020generative} to generate GUI prototypes based on GUI subtrees i.e. image-based excerpts extracted from \textit{Rico} GUIs. While the GUI prototypes generated by their approach possess a realistic appearance, they are image-based and therefore not interactive, which restricts the usefulness compared to the GUI prototypes generated by our approach. In addition, their approach requires extensive training resources. More recent approaches focus on employing LLMs for generating GUI prototypes of different fidelity. To generate low-fidelity GUI layouts based on high-level textual layout descriptions and a collection of selected GUI component types, \textit{Instigator} \cite{brie2023evaluating} trains a full \textit{minGPT} \cite{minGPT} model from scratch exploiting a large-scale repository of automatically scraped web pages, which are then transformed to low-fidelity layouts for training. Instead of training a task-specific GPT model from scratch, another approach has been proposed which fine-tunes a pretrained LLM for generating GUIs exploiting the \textit{Rico} GUI repository \cite{feng2023designing}. While both approaches generate GUI layouts or prototypes from brief text descriptions, they require resource-intensive training, create low-fidelity prototypes and generate merely domain-specific languages (DSL) to represent the GUI prototype, which limits the overall usefulness. Similarly, \textit{MAxPrototyper} \cite{yuan2024maxprototyper} generates a custom DSL to represent GUI prototypes from short text descriptions and a prespecified GUI layout using an LLM, however, their approach focuses mainly on generating GUI content in the form of text and images matching the provided text description. In addition, \textit{UIDiffuser} \cite{wei2023boosting} utilizes \textit{stable diffusion} \cite{rombach2022high} to directly generate GUI prototype images from short text descriptions. While this approach can provide coarse design inspirations, the generated images lack clarity and do not represent functioning GUI components. In addition, the generated format of the GUI prototypes in the form of images is difficult to integrate or adapt in GUI prototyping tools from practice.

\vspace{0.05cm}

\subsubsection{GUI Retrieval}
Many approaches have been proposed for NL-based GUI retrieval \cite{bernal2019guigle, kolthoff2019automatic, kolthoff2020gui2wire, kolthoff2021automated, huang2021creating, kolthoff2023data}. While these GUI retrieval methods facilitate GUI prototyping by rapidly retrieving relevant GUIs from large-scale GUI repositories, the returned GUIs are usually non-interactive and non-adaptable images. In particular, we integrated NL-based GUI retrieval with LLMs to generate interactive GUI prototypes. Moreover, we significantly enhance the prior GUI re-ranking performance by the proposed LLM-based re-ranking mechanism. In addition, approaches have been proposed that retrieve GUIs based on other input types, for example, \textit{GUIFetch} \cite{behrang2018guifetch} and \textit{Swire} \cite{huang2019swire} conducting sketch-based GUI retrieval by employing a deep neural network with nearest neighbour search, \textit{VINS}  \cite{bunian2021vins} conducting wireframe-based GUI retrieval by utilizing neural image and GUI component embedding models, \textit{Screen2Vec} \cite{li2021screen2vec} conducting GUI retrieval based on GUI screenshots by using a multi-modal neural GUI embedding model, \textit{Gallery DC} \cite{chen2019gallery} retrieving GUI components extracted from a large-scale GUI collection and \textit{d.tour} \cite{ritchie2011d} enabling stylistic keyword and template-based GUI searching.
\vspace{0.175cm}
\section{Conclusion \& Future Work}

In this paper, we proposed several ZS prompting approaches for enhancing GUI generation. Particularly, SCGG mostly significantly outperforms all other methods. Increasing the number of examples to a larger number in RAGG leads to increased performance over several metrics for GUI generation. However, RAGG would perform most effective when the LLM knowledge and GUI dataset are more complementary to each other. Including LLM-based content generation improves the effectiveness of generated GUI prototypes significantly.

While SCGG is an effective ZS prompting method to enhance generated GUI prototypes, the token consumption is large. For future work, we plan to increase the efficiency of SCGG by dynamically adapting the feedback type (e.g., focus only on design) and compute new prototypes in a decomposed manner i.e. updating only affected GUI sections.

\ifCLASSOPTIONcaptionsoff
  \newpage
\fi

\bibliographystyle{IEEEtran}
\bibliography{bibtex/bib/ref}

\end{document}